# Casimir-Polder interaction between an excited atom and a gas dielectric medium

Yury Sherkunov

Institute for High Energy Density of Joint Institute for High Temperatures, RAS,

Moscow, Russia

sherkunovyb@physics.org

The Casimir-Polder potential for interaction between an excited atom and a ground state one in the retarded case obtained with the help of perturbation technique drops as $R^{-2}$ with the distance between the atoms [E. A. Power and T. Thirunamachandran, Phys. Rev. A **47**, 2539 (1993)]. It results in divergent integrals for interaction between an excited atom and a dilute gas medium. We investigate interaction between two atoms embedded in a dielectric medium with the help of a method, which enables us to make calculations beyond the perturbation technique. We take into account absorption of photons in the medium. This approach solves the problem of divergence. We consider interaction between an excited atom and a planar dielectric gas medium of ground state atoms. We show that the retarded interaction between an excited atom and a gas of ground-state atoms is not oscillating but follows a simple power law. We show that to obtain conventional non-retarded expression for the van der Waals force between an excited atom and a dilute gas, the distance between the atom and the interface should be much smaller than the free mean path of a photon in the medium.

Interaction between an excited atom and a hemisphere of ground-state atoms is considered.

## I. Introduction

The dispersion interaction between two ground-state atoms separated by a distance large enough to neglect exchange interaction was studied for the first time by Fritz London in 1930[1]. Using quantum mechanics, he showed that the interaction potential between the atoms drops as $R^{-6}$ with the distance $R$ between the atoms. Later, Casimir[2] and Casimir and Polder[3] demonstrated, that for the distances larger than a transition wavelength of atoms $\lambda$, one should use quantum electrodynamics and take into account the electromagnetic vacuum to calculate the potential. According to Casimir and Polder, the interaction potential in the retardation regime is proportional to $R^{-7}$. The dispersion dipole-dipole interaction between two ground-state atoms is always attractive.

If one of the atoms is excited, the dispersion interaction differs significantly from the case of two ground-state atoms. It may be up to four orders of magnitude stronger than the interaction between two ground-state atoms. Moreover, the interaction may be either attractive, or repulsive depending on the transition frequencies of the atoms. Such interaction was studied both theoretically[4,5,6,7,8,9,10,11,12], using perturbation technique, and experimentally[13,14].

If the atoms are identical, one should describe them in terms of symmetric or antisimmetric eigenfunction. The corresponding eigenstates have a delocalized excitation (Frenkel exciton) in a symmetric or antisymmetric configuration over both atoms. This situation is not due to electron overlap but due to dipole-dipole interaction[15]. For non-retarded case ($R \ll \lambda$) the interaction potential is proportional to $R^{-3}$ [13,14,15], while for the retarded case ($R \gg \lambda$) the potential drops as $R^{-4}$ [10,11,15]. The interaction of identical atoms is studied in numerous papers theoretically and experimentally[10,11,13,14,15]. We will not consider this case here.

If the atoms are dissimilar and coupling energy is smaller than the splitting between the excited levels of the atoms, one can describe the atoms by independent eigenfunctions. In this case $U \sim R^{-6}$ for non-retarded regime and $U \sim R^{-2}$ for the retarded one. In the opposite case, one should consider the interaction as though the atoms were identical. The transition from non-degenerate (independent eigenfunctions) to degenerate (common eigenfunction) cases is discussed in [14].

Interaction between a ground-state atom and a dielectric or metal wall was studied for the first time by J.E. Lennard-Jones in 1932[16]. Since 1932, the problem of interaction between an atom and a surface has been discussed in numerous papers. For references see[17][18]. If the atom is not excited, the interaction is always attractive[19]. If the atom is excited, the interaction is either attractive or repulsive depending on its transition frequency [19][20][21]. This resonance character of interaction has been confirmed experimentally [17][22][23].

Let us analyze the results of perturbative approach to the interaction of dissimilar atoms if one of them is excited[8][9]. The result for the interaction energy was obtained with the help of the linear response theory [8] and quantum electrodynamics [9]. Both approaches resulted in the same formula for the excited atom A and the ground-state atom B separated by a distance R

$$U(R) = -\frac{1}{\pi}\int_0^\infty \alpha_A(iu)\alpha_B(iu)\frac{u^4}{R^2}\left[1+\frac{2}{uR}+\frac{5}{u^2R^2}+\frac{6}{u^3R^3}+\frac{3}{u^4R^4}\right]e^{-2uR}du$$
$$-\frac{4}{9}\frac{\left|d_{eg}^A\right|^2\left|d_{eg}^B\right|^2\omega_B\omega_A^4}{\left(\omega_B^2-\omega_A^2\right)R^2}\left[1+\frac{1}{\omega_A^2R^2}+\frac{3}{\omega_A^4R^4}\right].$$
(1)

We simplified the formula for the case of two-level atoms and took into account only electric-dipole interaction. Here $\omega_A$ and $\omega_B$ are the transition frequencies of excited atom A and the ground state atom B. $d_{eg}^A$ and $d_{eg}^B$ are the transition dipole moments of atoms A and B. $\alpha_A(iu)$ and $\alpha_B(iu)$ are the polarizabilities of atoms A and B. The second term of the equation (1) appears only if atom A is excited. It is interesting to consider the asymptotic limits of the formula (1). For small R separations (van der Waals limit) $(R \ll \lambda)$, where $\lambda$ is the wavelength corresponding to atom transitions, the result is proportional to $R^{-6}$. The contribution of the second term of the right-hand side of the equation (1) leads to

$$U(R) = -\frac{4}{3}\frac{\left|d_{eg}^A\right|^2\left|d_{eg}^B\right|^2\omega_B}{\left(\omega_B^2-\omega_A^2\right)R^6}.$$
(2)

For a large separation case $(R \gg \lambda)$ the expression (1) reads

$$U(R) = -\frac{4}{9} \frac{\left|d_{eg}^A\right|^2 \left|d_{eg}^B\right|^2 \omega_A^4 \omega_B}{\left(\omega_B^2 - \omega_A^2\right) R^2}. \tag{3}$$

We should stress here that the interaction potential (3) for excited and ground-state atoms is proportional to $R^{-2}$, while the potential for two ground-state atoms falls off as $R^{-7}$.

Now let us consider interaction of an excited atom with a semi-infinite medium of ground-state atoms. For the sake of simplicity we will consider a dilute gas of two-level atoms as a medium. Let the distance between the excited atom and the medium be $z_0 \gg \lambda$.

Taking into account only pair interactions, we can represent the energy as a volume integral over the semi-infinite medium

$$U_1(z_0) = \int dV U(R) n_0, \tag{4}$$

where $n_0$ is the density number of the atoms of the medium.

Substituting expression (3) into (4), we have

$$U_1(z_0) = -\frac{4}{9} \frac{\left|d_{eg}^A\right|^2 \left|d_{eg}^B\right|^2 \omega_A^4}{\left(\omega_B^2 - \omega_A^2\right)} n_0 \int \frac{dV}{R^2} \to \infty. \tag{5}$$

This integral is divergent.

Thus, we arrive at the conclusion that the perturbation theory resulting in the expression (3), can not be implemented to calculate the interaction energy between an excited atom and a semi-infinite medium of ground-state atoms for a dilute gas.

We will calculate the interaction energy of two dissimilar atoms when they are immersed in a dielectric medium if one of them is excited. To find the energy we will use the method offered in[24]. As we will show, the implementation of this method will result in a finite expression for the energy instead of formula(5).

This paper is composed in the following way.

In Section II we consider interaction between two dissimilar atoms embedded in a dielectric medium using the method based on the kinetic Green functions [24]. We took into account absorption of the photons in the medium. The result for the interaction potential is proportional to the exponent of imaginary part of the refractive index of the medium. So the problem of divergences is solved.

In Section III we apply the result obtained in Section III to interaction between an excited atom and a gas medium of ground-state atoms for the case of planer interface. We show that in the retarded case the resonance term does not oscillate but follows simple power low. We introduced a new parameter – free mean path of the photon in the $L_{ph}$. We showed that the well known results for interaction between an excited atom and a dielectric medium for non-retarded regime can be obtained only for $z_0 << L_{ph}$, $z_0 << \lambda$, where $z_0$ is the distance between the atom and the interface.

In Section IV we consider interaction between an excited atom and a hemisphere of dilute gas medium, if the atom is situated in the center of the hemisphere.

## II. Interaction between two atoms in a dielectric medium if one of them is excited.

We consider two dissimilar atoms A and B immersed in an absorbing dielectric medium. Let atom A be in the excited state and situated at a point with radius-vector $R_A$ and B in the ground state and situated at a point $R_B$. We suppose the electromagnetic field to be in its vacuum state. The exchange interaction is negligible. Let us suppose for the sake of simplicity that the radiation width of excited level of atom A is negligible in comparison with the width of the excited level of atom B.

The Hamiltonian of the system is as follows

$$\hat{H} = \hat{H}_A + \hat{H}_B + \hat{H}_{med} + \hat{H}_{ph} + \hat{H}_{int}, \quad (6)$$

where $\hat{H}_A = \sum_i \varepsilon^0_{Ai}\hat{b}^\dagger_i\hat{b}_i$, $\hat{H}_B = \sum_i \varepsilon^0_{Bi}\hat{\beta}^\dagger_i\hat{\beta}_i$, $\hat{H}_{med} = \sum_i \varepsilon^0_{medi}\hat{c}^\dagger_i\hat{c}_i$ are the Hamiltonians of noninteracting atoms A, B, and the atoms of the medium, $\varepsilon^0_i$ is the energy of i-th state of the corresponding atom without the Lamb shift, $\hat{b}_i(\hat{b}^\dagger_i), \hat{\beta}_i(\hat{\beta}^\dagger_i), \hat{c}_i(\hat{c}^\dagger_i)$ are annihilation (creation) operators of i-th state of corresponding atom, $\hat{H}_{ph} = \sum_{k\lambda} k\left(\hat{\alpha}^\dagger_{k\lambda}\hat{\alpha}_{k\lambda} + \frac{1}{2}\right)$ is the Hamiltonian of free electromagnetic field, $k$ is the wave vector, $\lambda = 1,2$ is the index of polarization of electromagnetic field, $\hat{\alpha}_{k\lambda}\left(\hat{\alpha}^\dagger_{k\lambda}\right)$ are annihilation (creation) operators of electromagnetic field.

The interaction Hamiltonian in the interaction picture is

$$\hat{H}_{int\,I}(t) = -\int \hat{\psi}^\dagger_l(x)\hat{d}^\nu \hat{E}^\nu_l(x)\hat{\psi}_l(x)d\mathbf{r} - \int \hat{\varphi}^\dagger_l(x)\hat{d}^\nu \hat{E}^\nu_l(x)\hat{\varphi}_l(x)d\mathbf{r} - \int \hat{\chi}^\dagger_l(x)\hat{d}^\nu \hat{E}^\nu_l(x)\hat{\chi}_l(x)d\mathbf{r} \quad (7)$$

where

$$\hat{\psi} = \sum_i \psi_i(\mathbf{r}-\mathbf{R}_A)e^{-i\varepsilon_{Ai}t}\hat{b}_i, \quad \hat{\varphi} = \sum_i \varphi_i(\mathbf{r}-\mathbf{R}_B)e^{-i\varepsilon_{Bi}t}\hat{\beta}_i, \quad \hat{\chi}_l(x) = \sum_i \chi_i(\mathbf{r}-\mathbf{R}_m)e^{-i\varepsilon_{medi}t}\hat{c}_i \quad (8)$$

with $\psi_i(\mathbf{r}-\mathbf{R}_A)$, $\varphi_i(\mathbf{r}-\mathbf{R}_B)$, and $\chi_i(\mathbf{r}-\mathbf{R}_m)$ being the wave functions of i-th state of corresponding atoms. $\hat{d}^\nu$ is the operator of dipole moment, $\hat{E}^\nu(\mathbf{r})$ is the operator of free electromagnetic field

$$\hat{E}^\nu(x) = i\sum_{k\lambda}\sqrt{\frac{2\pi k}{V}}e^\nu_{k\lambda}\left(\hat{\alpha}_{k\lambda}e^{i\mathbf{k}\mathbf{r}}e^{-i\omega\lambda(\lambda)t} - \hat{\alpha}^\dagger_{k\lambda}e^{-i\mathbf{k}\mathbf{r}}e^{i\omega(\lambda)t}\right), \quad (9)$$

where V is the quantization volume, $e^\nu_{k\lambda}$ is the polarization unit vector, $\mathbf{R}_m$ describes the position of an atom of the dielectric medium, $x = \{\mathbf{r},t\}$.

We will calculate the interaction potential of the atoms using the method of kinetic quantum Green's functions[25][26] applied to quantum electrodynamics[24]. This method enables us to handle the divergence of the integral (5). The interaction potential of the atoms can be expressed as the energy shift of one of the atoms (Appendix A)

$$U(\mathbf{R}_A - \mathbf{R}_B) = \Delta E_B. \tag{10}$$

We suppose that the Lamb shift as well as the shift due to the interaction with the atoms of the media excluding atom B is already taken into account in $\varepsilon_{A0}$, so we take into account only the interaction between atoms A and B.

Let

$$G^B_{ll'}(x, x') = -i\langle \hat{T}_c \tilde{\varphi}_l(x) \tilde{\varphi}^\dagger_{l'}(x') \rangle \tag{11}$$

be the Green function of atom B. Here $x = \{\mathbf{r}, t\}$, operators are in the Heisenberg representation, $\langle ... \rangle$ means averaging over initial state of free atoms and vacuum state of the electromagnetic field. $\hat{T}_c$ is the operator of time-ordering [25,26]. It acts as follows

$$\hat{T}_c \hat{A}_1(t) \hat{B}_1(t') = \begin{cases} \hat{A}(t)\hat{B}(t'), & t > t' \\ \hat{B}(t')\hat{A}(t), & t < t' \end{cases}, \quad \hat{T}_c \hat{A}_2(t) \hat{B}_2(t') = \begin{cases} \hat{B}(t')\hat{A}(t), & t > t' \\ \hat{A}(t)\hat{B}(t'), & t < t' \end{cases}, \tag{12}$$

$$\hat{T}_c \hat{A}_1(t) \hat{B}_2(t') = \hat{B}(t')\hat{A}(t), \quad \hat{T}_c \hat{A}_2(t) \hat{B}_1(t') = \hat{A}(t)\hat{B}(t').$$

If the exchange interaction can be omitted, the atoms can be considered as independent. Consequently we can describe each atom by its own matrix of density.

Using the Green function (11) it is easy to find the matrix of density of atom B

$$\rho^B(x, x') = iG^B_{12}(x, x'). \tag{13}$$

To find the Green function (11) we will use the interaction representation. An operator in the Heisenberg representation $\tilde{\varphi}$ is connected with the one in the interaction picture $\hat{\varphi}$ as follows[26]

$$\tilde{\varphi}_l(x) = \hat{S}^{-1}(t) \hat{\varphi}_l(x) \hat{S}(t), \tag{14}$$

where $\hat{S}(t)$ is the scattering operator. Substituting (14) into (11) results in[26]

$$G^B_{ll'}(x, x') = -i\langle \hat{T}_c \hat{S}^{-1}(-\infty, \infty) \hat{\varphi}_l(x) \hat{\varphi}^\dagger_{l'}(x') \hat{S}(-\infty, \infty) \rangle. \tag{15}$$

To simplify this formula we use the following generalized scattering operator[25][26]

$$\hat{S}_c = \hat{T}_c \exp\left\{\sum_{l=1,2}(-1)^l i \int_{-\infty}^{\infty} \hat{H}_{int\,l}(t)dt\right\} \qquad (16)$$

Thus, Eq. (11) can be rewritten

$$G^B_{ll'}(x,x') = -i\langle \hat{T}_c \hat{\varphi}_l(x)\hat{\varphi}^\dagger_{l'}(x')\hat{S}_c\rangle. \qquad (17)$$

To find the expression for the density matrix (13) we will represent the matrix of scattering (16) as a perturbation expansion and substitute it into Eq.(17).

The first two orders of the perturbation theory give

$$\rho^B(x,x') = \rho^B_0(x,x')$$
$$-\frac{1}{2}\langle \hat{T}_c \int dx_1 dx_2 \hat{\varphi}_1(x)\hat{\varphi}^\dagger_2(x')(-1)^{l_1+l_2}\hat{\varphi}^\dagger_{l_1}(x_1)\hat{E}^{v_1}_{l_1}(x_1)\hat{d}^{v_1}\hat{\varphi}_{l_1}(x_1)\hat{\varphi}^\dagger_{l_2}(x_2)\hat{E}^{v_2}_{l_2}(x_2)\hat{d}^{v_2}\hat{\varphi}_{l_2}(x_2)\rangle.$$

We describe atoms A and B in terms of independent initial eigenfunctions. Consequently, the normal ordering of operators describing atom B is

$$\langle \hat{N}\hat{\varphi}_{l_1}(x)\hat{\varphi}^\dagger_{l_2}(x')\hat{\varphi}^\dagger_{l'}(x_1)\hat{\varphi}_{l'}(x_2)...\rangle = 0,$$

for any order but the second one, while the second order of normal product represents the density matrix of initial state of atom

$$\rho^B_0(x,x') = \langle \hat{N}\hat{\varphi}_{l_1}(x)\hat{\varphi}^\dagger_{l_2}(x')\rangle = \langle \hat{\varphi}^\dagger(x')\hat{\varphi}(x)\rangle.$$

In the energy domain this matrix of density is

$$\rho^B_0(E,\mathbf{r}-\mathbf{r}') = 2\pi\sum_i \varphi_i(\mathbf{r}-\mathbf{R}_B)\varphi^*_i(\mathbf{r}'-\mathbf{R}_B)\delta(E-\varepsilon_{Bi})\langle \hat{\beta}^\dagger_i \hat{\beta}_i\rangle.$$

If the initial state of atom B is ground state, then

$$\rho^B_0(E,\mathbf{r}-\mathbf{r}') = 2\pi\varphi_g(\mathbf{r}-\mathbf{R}_B)\varphi^*_g(\mathbf{r}'-\mathbf{R}_B)\delta(E-\varepsilon_{Bg}). \qquad (18)$$

Index g stands for ground state.

Using Wick's theorem, we find

$$\rho^B(x,x') = \rho_0^B(x,x')$$
$$-i\int dx_1 dx_2 \rho_0^B(x,x_1)\hat{d}^v\hat{d}^{v'}g_{22}^{0B}(x_1,x_2)D_{22}^{0vv'}(x_2,x_1)g_{22}^{0B}(x_2,x')$$
$$-i\int dx_1 dx_2 g_{11}^{0B}(x,x_1)\hat{d}^v\hat{d}^{v'}g_{11}^{0B}(x_1,x_2)D_{11}^{0vv'}(x_2,x_1)\rho_0^B(x_2,x') \quad (19)$$
$$-i\int dx_1 dx_2 g_{11}^{0B}(x,x_1)\hat{d}^v\hat{d}^{v'}\rho_0^B(x_1,x_2)D_{21}^{0vv'}(x_2,x_1)g_{22}^{0B}(x_2,x').$$

The Feynman diagrams corresponding to Eqs.(19) are given in Fig.1. The first term is given in Fig.1(a), the second term is given in Fig1(b), the therd term is given in Fig.1(c), and the fourth term representing the incoherent channel is shown in Fig.1(d).

The propagators of free atom B are

$$g_{11}^{0B}(x,x') = -i\langle \hat{T}_c\hat{\varphi}_1(x)\hat{\varphi}_1^\dagger(x')\rangle_{vac} = -i\theta(t-t')\sum_i \varphi_i^*(r')\varphi_i(r)e^{-i\varepsilon_{Bi}(t-t')},$$
$$g_{22}^{0B}(x,x') = -i\langle \hat{T}_c\hat{\varphi}_2(x)\hat{\varphi}_2^\dagger(x')\rangle_{vac} = -i\theta(t'-t)\sum_i \varphi_i^*(r')\varphi_i(r)e^{-i\varepsilon_{Bi}(t-t')},$$
$$g_{12}^{0B}(x,x') = -i\langle \hat{\varphi}_2^\dagger(x')\hat{\varphi}_1(x)\rangle_{vac} = 0, \quad (20)$$
$$g_{21}^{0B}(x,x') = -i\langle \hat{\varphi}_2(x)\hat{\varphi}_1^\dagger(x')\rangle_{vac} = -i\sum_i \varphi_i^*(r')\varphi_i(r)e^{-i\varepsilon_{Bi}(t-t')},$$

where $\theta(t-t')$ is the unit step-function.

Deriving Eqs. (19), we took into account $g_{12}^{0B} = 0$.

The free photon propagator is

$$D_{ll'}^{0vv'}(x',x) = i\langle \hat{T}_c\hat{E}_l^v(x')\hat{E}_{l'}^{v'}(x)\rangle_{vac}, \quad (21)$$

The first three terms of the Eq. (19) correspond to the so-called coherent channel ($\rho_c^B$), with atom B returning to its initial state (e.g. elastic scattering). The last term represents the so-called incoherent channel ($\rho_n^B$), with atom B changing its initial state (e.g. spontaneous decay, Raman scattering). Consequently, we can represent the matrix of density as a sum

$$\rho^B(x,x') = \rho_c^B(x,x') + \rho_n^B(x,x').$$

The representation of the density matrix as a sum of contributions of the coherent channel and the incoherent one was used by B. Veklenko[27] for electromagnetic field and [28] for a group of atoms.

Here, we are interested in the Casimir-Polder interaction only. Consequently, atom B does not change its initial state as a result of interaction. It means that the incoherent channel does not contribute to the interaction potential.

The simplest Feynman diagrams for the fourth-order perturbation expression for the coherent channel are given in Fig.2. Here we did not take into account interaction between atoms A and B. The expression given by the diagram in Fig. 2(a) can be expressed via the diagram given in Fig.1(b), as well as the diagram in Fig.2(b) can be expressed via diagram in Fig.1(c).

The diagram in Fig.2(c) corresponds to the expression

$$\rho_c^{(4)B}(x,x') = \int dx_1 dx_2 dx_3 dx_4 g_{11}^{0B}(x,x_1) \hat{d}^v \hat{d}^{v'} g_{11}^{0B}(x_1,x_2) D_{11}^{0vv'}(x_2,x_1)$$
$$\times \rho_0^B(x_2,x_3) \hat{d}^{v_1} \hat{d}^{v_2} g_{22}^{0B}(x_3,x_4) D_{22}^{0v_1 v_2}(x_4,x_3) g_{22}^{0B}(x_4,x').$$

To demonstrate the method of summation of Feynman's diagrams, let us consider diagrams in Fig.1(b) and 2(a). If we calculate all the sequence of the diagrams of the same type, we will find the diagram in Fig.3. Thick solid line represents the complete propagators $g_{11}^B(x,x')\left(g_{22}^B(x,x')\right)$ of atom B, which obey the Dyson equations[29]

$$g_{11}^B(x,x') = g_{11}^{0B}(x,x') + \int dx_1 dx_2 g_{11}^{0B}(x,x_1) M_{11}(x_1,x_2) g_{11}^B(x_2,x'),$$
$$g_{22}^B(x,x') = g_{22}^{0B}(x,x') + \int dx_1 dx_2 g_{22}^{0B}(x,x_1) M_{22}(x_1,x_2) g_{22}^B(x_2,x').$$
(22)

$M_{11}$ and $M_{22}$ are the mass operators given by the following equations

$$M_{11}(x,x') = -i\hat{d}^v \hat{d}^{v'} g_{11}^B(x,x') D_{11}^{0vv'}(x',x),$$
$$M_{22}(x,x') = i\hat{d}^v \hat{d}^{v'} g_{22}^B(x,x') D_{22}^{0vv'}(x',x).$$
(23)

The real parts of mass operators (23) describe the Lamb shift, while the imaginary parts describe the radiation widths of excited levels of atoms.

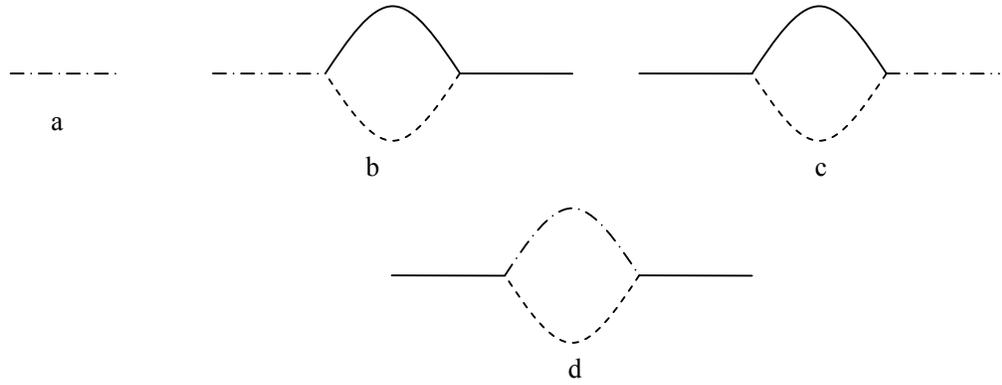

**Fig.1. Feynman's diagrams for the lowest orders of perturbation theory. Solid line corresponds to atom propagator $g^{0B}$. Dashed line corresponds to photon propagator $D^0$. Dashed-dotted line represents density matrix $\rho_B^0$. The coherent channel is represented by diagrams (a)-(c), the incoherent channel is represented by diagram (d).**

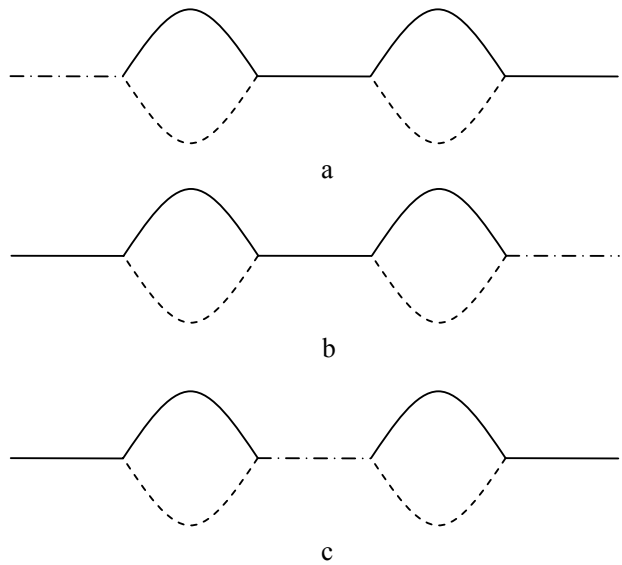

**Fig.2. Feynman's diagrams for the simplest expressions of the fourth order perturbation technique.**

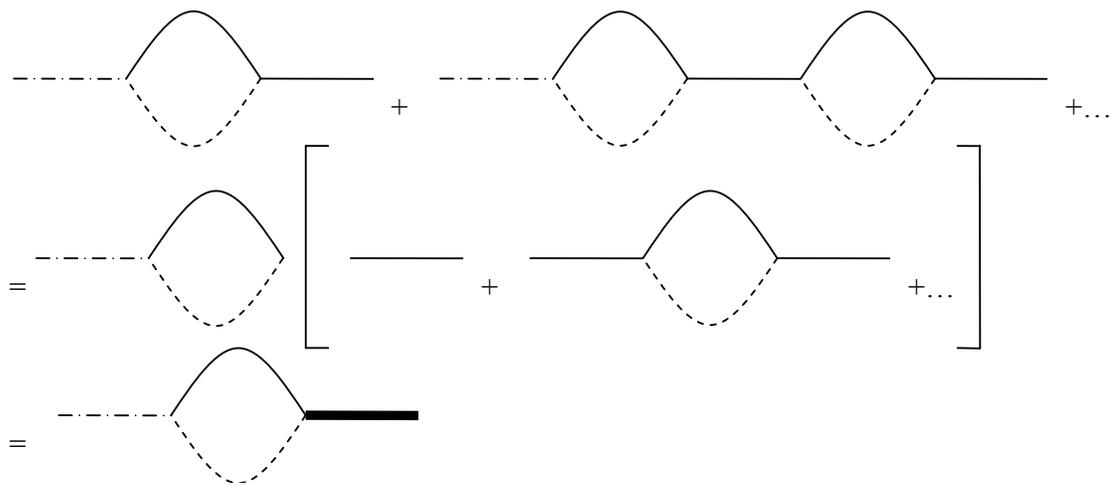

**Fig.3 Summation of Feynman's diagrams**

To calculate the Green function $g_{11}^B$ of atom B taking into account the Lamb shift and radiation line width of atom B we should substitute Eq.(23) into (22) and find the Fourier transformation. After renormalization, we find

$$g_{11}^B(E,\boldsymbol{r},\boldsymbol{r'}) = \sum_i \frac{\varphi_i(\boldsymbol{r}-\boldsymbol{R}_B)\varphi_i^*(\boldsymbol{r'}-\boldsymbol{R}_B)}{E-\varepsilon_{Bi}+i\frac{\gamma_{Bi}}{2}}. \qquad (24)$$

For atom A we neglect the width of excited state energy, consequently,

$$g_{11}^A(E,\boldsymbol{r},\boldsymbol{r'}) = \sum_i \frac{\psi_i(\boldsymbol{r}-\boldsymbol{R}_A)\psi_i^*(\boldsymbol{r'}-\boldsymbol{R}_A)}{E-\varepsilon_{Ai}+i0}. \qquad (25)$$

Now, summing up Feynman's diagrams, we find

$$\begin{aligned}\rho_c^B(x,x') &= \rho_0^B(x,x') + \int dx_1 dx_2 g_{11}^B(x,x_1) M_{11}(x_1,x_2)\rho_0^B(x_2,x') \\ &+ \int dx_1 dx_2 \rho_0^B(x,x_1) M_{22}(x_1,x_2) g_{22}^B(x_2,x') \\ &+ \int dx_1 dx_2 dx_3 dx_4 g_{11}^B(x,x_1) M_{11}(x_1,x_2)\rho_0^B(x_2,x_3) M_{22}(x_3,x_4) g_{22}^B(x_4,x'),\end{aligned} \qquad (26)$$

with the mass operators given by (23).

Eq. (25) along with the Eq. (23) describes the interaction of atom B with the vacuum, which results in the Lamb shift and radiation line-broadening. The interaction with atom A and the medium could be taken into account the similar way[30]. We should consider higher orders of the expansion of the scattering operator (16). Finally, we come to the Eq. (26) with the mass operator given by (Appendix B)

$$\begin{aligned}M_{11}(x,x') &= -i\hat{d}^\nu \hat{d}^{\nu'} g_{11}^B(x,x') D_{11}^{\nu\nu'}(x',x), \\ M_{22}(x,x') &= i\hat{d}^\nu \hat{d}^{\nu'} g_{22}^B(x,x') D_{22}^{\nu\nu'}(x',x).\end{aligned} \qquad (27)$$

Here $D_{11}^{\nu\nu'}(x',x)$ and $D_{22}^{\nu\nu'}(x',x)$ are the complete photon Green functions with account of the interaction with atoms.

$$D_{ll'}^{\nu\nu'}(x',x) = i\langle \hat{T}_c \tilde{E}_l^\nu(x') \tilde{E}_{l'}^{\nu'}(x)\rangle_{vac} = i\langle \hat{T}_c \hat{E}_l^\nu(x') \hat{E}_{l'}^{\nu'}(x) \hat{S}_c\rangle_{vac}, \qquad (28)$$

$\tilde{E}$ is in the Heisenberg representation and $\hat{E}$ is in the interaction representation.

So, Eq. (26) along with the Eq.(27) take into account not only the interaction of atom B with the vacuum (Lamb shift), but the interaction with atom A and the medium.

The integral equation (26) can be rewritten as a differential one [24].

$$\rho_c^B(x,x") = \Psi(x)\Psi^*(x"),$$
$$\left(i\frac{\partial}{\partial t} - \hat{H}_B\right)\Psi(x) = \int M_{11}(x,x')\Psi(x')dx', \tag{29}$$

The coherent channel processes do not change the initial state of atom B, consequently

$$\Psi(x) = \varphi_0(\mathbf{r} - \mathbf{R}_B)f(t), \tag{30}$$

where index "0" stands for initial state of atom B. Substituting (30) into (29) and neglecting non-diagonal elements of the mass operator we arrive at the following equation

$$i\frac{\partial}{\partial t}f(t) - \varepsilon_{0A}f(t) = \int_{t_0}^{\infty}\langle M_{11}(t,t')\rangle f(t')dt',$$

$$\langle M_{11}(t,t_1)\rangle = \int \varphi_0^*(\mathbf{r} - \mathbf{R}_B)M_{11}(x,x_1)\varphi_0(\mathbf{r}_1 - \mathbf{R}_B)d\mathbf{r}d\mathbf{r}_1, \tag{31}$$

here we suppose that the interaction was switched on at $t_0$ ($t_0 \to -\infty$).

Using pole approximation we find

$$\Psi(x) = \varphi_0(\mathbf{r} - \mathbf{R}_B)e^{-i\varepsilon_{B0}t}e^{-i\langle M_{11}(\varepsilon_{B0})\rangle(t-t_0)},$$

where $\langle M_{11}(\varepsilon_{B0})\rangle = \int_{-\infty}^{\infty}\langle M_{11}(t,t')\rangle e^{i\varepsilon_{B0}(t-t')}d(t-t')$ is the Fourier transform of mass operator taken at point $E = \varepsilon_{B0}$.

Thus, the density matrix of coherent channel in energy domain

$$\rho_c^B(E,E',\mathbf{r},\mathbf{r}') = \int_{t_0}^{\infty}\rho_c^B(x,x')e^{iEt-iE't'}dtdt', t_0 \to -\infty$$

is

$$\rho_c^B(E,E',\mathbf{r},\mathbf{r}') = \frac{\varphi_0(\mathbf{r} - \mathbf{R}_B)\varphi_0^*(\mathbf{r}' - \mathbf{R}_B)e^{i(E-E')t_0}}{\left(E - \varepsilon_{B0} - \langle M_{11}(\varepsilon_{B0})\rangle\right)\left(E' - \varepsilon_{B0} - \langle M_{22}(\varepsilon_{B0})\rangle\right)}. \tag{32}$$

Consequently, the interaction potential is expressed in terms of the mass operator $M_{11}^{00}$

$$U(\mathbf{R}_A - \mathbf{R}_B) = \Delta E_B = \text{Re}\langle M_{11}(\varepsilon_{B0})\rangle. \tag{33}$$

Thus, to calculate the interaction potential we have to find the mass operator (27) and the photon Green functions (28). Substituting (27) into (33), we find the general relation between the interaction potential and the Green function of the photons.

$$U(\mathbf{R}_A - \mathbf{R}_B) = -\text{Re}\left[\int_{-\infty}^{\infty} d\omega \int d\mathbf{r} d\mathbf{r}' \varphi_0^*(\mathbf{r} - \mathbf{R}_B) \frac{i\hat{d}^\nu \hat{d}^{\nu'}}{2\pi} g_{11}^B(\omega + \varepsilon_{B0}, \mathbf{r}, \mathbf{r}') D_{11}^{\nu\nu'}(\omega, \mathbf{r}', \mathbf{r}) \varphi_0(\mathbf{r}' - \mathbf{R}_B)\right]. \tag{34}$$

Now, we substitute Eq.(24) into Eq.(34) and calculate the integrals using the following expressions for the matrix elements of the dipole moments in dipole approximation

$$d_{mn}^{B\nu} e^{i\mathbf{k}\mathbf{R}_B} = \int \varphi_m^*(\mathbf{r} - \mathbf{R}_B) \hat{d}^\nu e^{i\mathbf{k}\mathbf{r}} \varphi_n(\mathbf{r} - \mathbf{R}_B) d\mathbf{r}. \tag{35}$$

$$U(\mathbf{R}_A - \mathbf{R}_B) = -\text{Re}\frac{id_{eg}^{B\nu} d_{eg}^{*B\nu'}}{2\pi} \int_{-\infty}^{\infty} d\omega \frac{D_{11}^{\nu\nu'}(\omega, \mathbf{R}_B, \mathbf{R}_B)}{\omega - \omega_B + i\frac{\gamma_B}{2}}. \tag{36}$$

Here $\omega_B$ is the transition frequency of atom B, $d_{eg}^{B\nu}$ is the matrix element of dipole moment. Index g stands for ground state and index e stands for excited state.

The Eq. (36) corresponds to the interaction between an atom B and electromagnetic field described by the Green function $D_{11}^{\nu\nu'}$ at zero temperatures. So, we can investigate interaction between atom B and a body of arbitrary shape, provided the photon Green function is know.

Using Eqs. (28) and (12), we can write

$$D_{11}^{vv'}(x',x) = i\langle \tilde{E}^v(x')\tilde{E}^{v'}(x)\rangle_{vac}\theta(t'-t) + i\langle \tilde{E}^{v'}(x)\tilde{E}^v(x')\rangle_{vac}\theta(t-t'),$$

$$D_{12}^{vv'}(x',x) = i\langle \tilde{E}^{v'}(x)\tilde{E}^v(x')\rangle_{vac},$$

$$D_{21}^{vv'}(x',x) = i\langle \tilde{E}^v(x')\tilde{E}^{v'}(x)\rangle_{vac}, \tag{37}$$

$$D_{22}^{vv'}(x',x) = i\langle \tilde{E}^{v'}(x)\tilde{E}^v(x')\rangle_{vac}\theta(t'-t) + i\langle \tilde{E}^v(x')\tilde{E}^{v'}(x)\rangle_{vac}\theta(t-t').$$

These functions are connected with the retarded and advanced Green functions[31]

$$D_r^{vv'}(x',x) = i\langle \tilde{E}^v(x')\tilde{E}^{v'}(x) - \tilde{E}^{v'}(x)\tilde{E}^v(x')\rangle_{vac}\theta(t'-t),$$

$$D_a^{vv'}(x',x) = -i\langle \tilde{E}^v(x')\tilde{E}^{v'}(x) - \tilde{E}^{v'}(x)\tilde{E}^v(x')\rangle_{vac}\theta(t-t'). \tag{38}$$

Comparing Eqs. (37) and (38), we find [26][27]

$$D_r = D_{11} - D_{12} = D_{21} - D_{22},$$
$$D_a = D_{11} - D_{21} = D_{12} - D_{22}. \tag{39}$$

Using Eq. (37), we can establish another property of photon Green function in energy-coordinate domain

$$D_{11}^{vv'}(\omega,\mathbf{r},\mathbf{r}') = D_{11}^{vv'}(-\omega,\mathbf{r}',\mathbf{r}). \tag{40}$$

Using (39) and (40), we can rewrite (36).

$$U(\mathbf{R}_A - \mathbf{R}_B)$$
$$= \mathrm{Re}\frac{id_{eg}^{Bv}d_{eg}^{*Bv'}}{2\pi}\int_0^\infty d\omega\left(D_r^{vv'}(\omega,\mathbf{R}_B,\mathbf{R}_B) + D_{12}^{vv'}(\omega,\mathbf{R}_B,\mathbf{R}_B)\right)\left(\frac{1}{\omega_B - \omega - i\frac{\gamma_B}{2}} + \frac{1}{\omega_B + \omega - i\frac{\gamma_B}{2}}\right). \tag{41}$$

Direct calculation of the retarded Green function results in[31]

$$D_r^{vv'}(x',x) = D_r^{0vv'}(x',x) + \int dx_1 dx_2 \sum_{v_1 v_2} D_r^{0vv_1}(x',x_1)\Pi_r^{v_1 v_2}(x_1,x_2)D_r^{0v_2 v'}(x_2,x)$$
$$\int dx_1 dx_2 dx_3 dx_4 \sum_{v_1 v_2} D_r^{0vv_1}(x',x_1)\Pi_r^{v_1 v_2}(x_1,x_2)D_r^{0v_2 v'}(x_2,x_3)\Pi_r^{v_1 v_2}(x_3,x_4)D_r^{0v_2 v'}(x_4,x) + \ldots \tag{42}$$

Eq. (42) can be rewritten as the Dyson integral equation

$$D_r^{vv'}(x',x) = D_r^{0vv'}(x',x) + \int dx_1 dx_2 \sum_{v_1 v_2} D_r^{0vv_1}(x',x_1)\Pi_r^{v_1 v_2}(x_1,x_2)D_r^{v_2 v'}(x_2,x). \tag{43}$$

For the advanced Green function

$$D_a^{vv'}(x',x) = D_a^{0vv'}(x',x) + \int dx_1 dx_2 \sum_{v_1 v_2} D_a^{0vv_1}(x',x_1)\Pi_a^{v_1 v_2}(x_1,x_2)D_a^{v_2 v'}(x_2,x). \tag{44}$$

We use the retarded and advanced polarization operators[26][27]

$$\Pi_r = \Pi_{11} + \Pi_{12} = -(\Pi_{22} + \Pi_{21}),$$
$$\Pi_a = \Pi_{11} + \Pi_{21} = -(\Pi_{22} + \Pi_{12}).$$
(45)

Now, let us consider electromagnetic field in infinite medium described by the permittivity $\varepsilon(\omega)$.

Then the equation (43) can be rewritten in terms of the polarization operator of the medium as

$$D_{rmed}^{vv'}(x',x) = D_r^{0vv'}(x',x) + \int dx_1 dx_2 \sum_{v_1 v_2} D_r^{0vv_1}(x',x_1) \Pi_{rmed}^{v_1 v_2}(x_1,x_2) D_{rmed}^{v_2 v'}(x_2,x). \quad (46)$$

On the other hand the retarded Green function of photons obeys the equation[31]

$$\left( \frac{\partial^2}{\partial r^v \partial r^{v'}} - \delta_{vv'} \Delta - \omega^2 \varepsilon(\omega) \delta_{vv'} \right) D_{rmed}^{vv'}(r,r',\omega) = 4\pi \omega^2 \delta_{vv'} \delta(r-r'). \quad (47)$$

As a result, we arrive at a well known relation[31]

$$\Pi_{rmed}^{vv'}(k,\omega) = \int \exp(-ik(r-r') + i\omega(t-t')) \Pi_{rmed}^{vv'}(r-r',t-t') d(r-r') d(t-t')$$
$$= \frac{\delta_{vv'}(\varepsilon(\omega)-1)\omega^2}{4\pi}.$$
(48)

Eq. (48) enables us to calculate the permittivity of the medium, using expression (87) and high-order corrections to (87).

For the retarded Green function of the photons in $(k,\omega)$ domain in infinite medium we find the equation using (46) and (48)

$$D_{rmed}^{vv'}(k,\omega) = -4\pi\omega^2 \left( \delta_{vv'} - \frac{k_v k_{v'}}{\varepsilon(\omega)\omega^2} \right) \left( \varepsilon(\omega)\omega^2 - k^2 \right)^{-1}. \quad (49)$$

In the coordinate-energy domain we have

$$D_{rmed}^{vv'}(\omega, r-r') = \frac{1}{(2\pi)^3} \int_{-\infty}^{\infty} D_{rmed}^{vv'}(\omega, k) \exp(ik(r-r')) d(r-r')$$
$$= \omega^2 \left[ \delta_{vv'} \left( 1 + \frac{i}{n(\omega)\omega|r-r'|} - \frac{1}{n^2(\omega)\omega^2|r-r'|^2} \right) \right.$$
$$\left. + \frac{(r-r')_v (r-r')_{v'}}{|r-r'|^2} \left( \frac{3}{n^2(\omega)\omega^2|r-r'|^2} - \frac{3i}{n(\omega)\omega|r-r'|} - 1 \right) \right] \frac{e^{in(\omega)\omega|r-r'|}}{|r-r'|},$$
$$D_a^{vv'}(\omega, r-r') = \left( D_r^{vv'}(\omega, r-r') \right)^*$$
(50)

where $n(\omega) = \sqrt{\varepsilon(\omega)}$ is the complex refractive index of the medium.

Now, let us tern to our problem of interaction between two atoms embedded in a dielectric.

We use the Eq.(43) and denote

$$\Pi_A = \Pi - \Pi_{med} \tag{51}$$

the polarization operator of atom A only.

$$\begin{aligned}D_r &= D_r^0 + D_r^0 \Pi_{rmed} D_r^0 + D_r^0 \Pi_{rmed} D_r^0 \Pi_{rmed} D_r^0 + D_r^0 \Pi_{rmed} D_r^0 \Pi_{rmed} D_r^0 \Pi_{rmed} D_r^0 + ... \\ &+ D_r^0 \Pi_{Ar} D_r^0 + D_r^0 \Pi_{rmed} D_r^0 \Pi_{Ar} D_r^0 + D_r^0 \Pi_{rmed} D_r^0 \Pi_{rmed} D_r^0 \Pi_{Ar} D_r^0 + ...\end{aligned} \tag{52}$$

Here, we omit integrals for the sake of simplicity. The first line of (52) can be rewritten with the help of (46)

$$D_r^0 + D_r^0 \Pi_{rmed} D_r^0 + D_r^0 \Pi_{rmed} D_r^0 \Pi_{rmed} D_r^0 + D_r^0 \Pi_{rmed} D_r^0 \Pi_{rmed} D_r^0 \Pi_{rmed} D_r^0 + ... = D_{rmed},$$

while the second line reads

$$D_r^0 \Pi_{Ar} D_r^0 + D_r^0 \Pi_{rmed} D_r^0 \Pi_{Ar} D_r^0 + D_r^0 \Pi_{rmed} D_r^0 \Pi_{rmed} D_r^0 \Pi_{Ar} D_r^0 + ... = D_{rmed} \Pi_{Ar} D_r^0.$$

So, the Eq. (52) is

$$\begin{aligned}D_r &= D_{rmed} + D_{rmed} \Pi_{Ar} D_r^0 + D_{rmed} \Pi_{Ar} D_r^0 \Pi_{rmed} D_r^0 + D_{rmed} \Pi_{Ar} D_r^0 \Pi_{rmed} D_r^0 \Pi_{rmed} D_r^0 + ... \\ &= D_{rmed} + D_{rmed} \Pi_{Ar} D_{rmed} + ... = D_{rmed} + D_{rmed} \Pi_{Ar} D_{rmed} + D_{rmed} \Pi_{Ar} D_{rmed} \Pi_{Ar} D_{rmed} + ...\end{aligned}$$

Finally, we come to the Dyson equation

$$D_r^{vv'}(x',x) = D_{rmed}^{vv'}(x',x) + \int dx_1 dx_2 \sum_{v_1 v_2} D_{rmed}^{vv_1}(x',x_1) \Pi_{Ar}^{v_1 v_2}(x_1,x_2) D_r^{v_2 v'}(x_2,x). \tag{53}$$

Or in the first approximation

$$D_r^{vv'}(x',x) = D_{rmed}^{vv'}(x',x) + \int dx_1 dx_2 \sum_{v_1 v_2} D_{rmed}^{vv_1}(x',x_1) \Pi_{Ar}^{v_1 v_2}(x_1,x_2) D_{rmed}^{v_2 v'}(x_2,x). \tag{54}$$

It reminds the equation (43), but instead of the free photon Green function $D^0$ we have the Green functions of the photons in absorbing medium $D_{med}$.

The same type of consideration results in the following equation[27] for $D_{12}$.

$$\begin{aligned}D_{12}^{vv'}(x',x) = &\, D_{12med}^{vv'}(x',x) + \int dx_1 dx_2 \sum_{v_1 v_2} D_{rmed}^{vv_1}(x',x_1) \Pi_{Ar}^{v_1 v_2}(x_1,x_2) D_{12med}^{v_2 v'}(x_2,x) \\ &+ \int dx_1 dx_2 \sum_{v_1 v_2} D_{12med}^{vv_1}(x',x_1) \Pi_{Aa}^{v_1 v_2}(x_1,x_2) D_{amed}^{v_2 v'}(x_2,x) \\ &- \int dx_1 dx_2 \sum_{v_1 v_2} D_{rmed}^{vv_1}(x',x_1) \Pi_{A12}^{v_1 v_2}(x_1,x_2) D_{amed}^{v_2 v'}(x_2,x)\end{aligned} \tag{55}$$

For the Green functions $D_{12}^0$ and we can write the equations using (37)

$$D_{12}^{0vv'}(\omega, \mathbf{r}-\mathbf{r}') = \frac{2i\omega^2 \theta(-\omega)}{|\mathbf{r}-\mathbf{r}'|}\left[\delta_{vv'}\left(\sin(|\omega||\mathbf{r}-\mathbf{r}'|)\left(1-\frac{1}{\omega^2|\mathbf{r}-\mathbf{r}'|^2}\right)+\frac{\cos(\omega|\mathbf{r}-\mathbf{r}'|)}{|\omega||\mathbf{r}-\mathbf{r}'|}\right)\right.$$
$$\left.+\frac{(\mathbf{r}-\mathbf{r}')_v(\mathbf{r}-\mathbf{r}')_{v'}}{|\mathbf{r}-\mathbf{r}'|^2}\left(\sin(|\omega||\mathbf{r}-\mathbf{r}'|)\left(\frac{3}{\omega^2|\mathbf{r}-\mathbf{r}'|^2}-1\right)-\frac{3\cos(\omega|\mathbf{r}-\mathbf{r}'|)}{|\omega||\mathbf{r}-\mathbf{r}'|}\right)\right]. \tag{56}$$

We should stress here, that $D_{12}^{0vv'} = 0$ for $\omega > 0$. The situation does not change in the medium for it does not emit radiation[27]

$$D_{12med}^{vv'} = 0 \text{ for } \omega > 0. \tag{57}$$

To calculate the interaction potential, we should substitute (54) and (55) into (36) and take into account (57)

$$U(\mathbf{R}_A - \mathbf{R}_B) = \text{Re}\frac{id_{eg}^{Bv}d_{eg}^{*Bv'}}{2\pi}\int_0^\infty d\omega \left(\frac{1}{\omega_B - \omega - i\frac{\gamma_B}{2}} + \frac{1}{\omega_B + \omega - i\frac{\gamma_B}{2}}\right)$$
$$\times\left[D_{rmed}^{vv'}(\omega, \mathbf{R}_B, \mathbf{R}_B) + \sum_{v_1 v_2} D_{rmed}^{vv_1}(\omega, \mathbf{R}_B, \mathbf{R}_A)\Pi_{Ar}^{v_1 v_2}(\omega)D_{rmed}^{v_2 v'}(\omega, \mathbf{R}_B, \mathbf{R}_A)\right. \tag{58}$$
$$\left. - D_{rmed}^{vv_1}(\omega, \mathbf{R}_B, \mathbf{R}_A)\Pi_{A12}^{v_1 v_2}(\omega)D_{amed}^{v_2 v'}(\omega, \mathbf{R}_B, \mathbf{R}_A)\right].$$

The first term in the brackets corresponds to the interaction of atom B with the medium and vacuum. Account of this term result in substitution $\varepsilon_B^0 \to \varepsilon_B$, with $\varepsilon_B^0$ being the energy of bare state of atom B, and $\varepsilon_B$ being the energy of the state with account of the Lamb shift and interaction with the medium. Direct calculation of the polarization operators with the help of (86), (51), and (45) gives us

$$\Pi_{Ar}^{v_1 v_2}(\omega) = -d_{eg}^{*Av_1}d_{eg}^{Av_2}\left(\frac{1}{\omega_A + \omega + i0} + \frac{1}{\omega_A - \omega - i0}\right) \tag{59}$$

$$\Pi_{A12}^{v_1 v_2}(\omega) = -2\pi i d_{eg}^{*Av_1}d_{eg}^{Av_2}\delta(\omega - \omega_A) \tag{60}$$

The polarization operator is equal to the polarizability of atom A.

$$\Pi_{Ar}^{v_1 v_2}(\omega) = \alpha_A^{v_1 v_2}(\omega),$$

with the polarizabilities of excited and ground-state atoms being[29]

$$\alpha_g^{vv'}(\omega) = \frac{d_{ge}^v d_{eg}^{v'}}{\omega_{eg} - \omega - i\frac{\gamma}{2}} + \frac{d_{eg}^v d_{ge}^{v'}}{\omega_{eg} + \omega + i\frac{\gamma}{2}}, \tag{61}$$

$$\alpha_e^{vv'}(\omega) = \frac{d_{eg}^v d_{ge}^{v'}}{-\omega_{eg} - \omega - i\frac{\gamma}{2}} + \frac{d_{ge}^v d_{eg}^{v'}}{-\omega_{eg} + \omega + i\frac{\gamma}{2}} \tag{62}$$

Thus, we find

$$U(\mathbf{R}_A - \mathbf{R}_B) = \mathrm{Re}\frac{id_{eg}^{Bv} d_{eg}^{*Bv'}}{2\pi} \int_0^\infty d\omega \left( \frac{1}{\omega_B - \omega - i\frac{\gamma_B}{2}} + \frac{1}{\omega_B + \omega - i\frac{\gamma_B}{2}} \right)$$
$$\times \sum_{v_1 v_2} \left[ D_{rmed}^{vv_1}(\omega, \mathbf{R}_B, \mathbf{R}_A) \alpha_A^{v_1 v_2} D_{rmed}^{v_2 v'}(\omega, \mathbf{R}_B, \mathbf{R}_A) \right. \tag{63}$$
$$\left. + 2\pi i D_{rmed}^{vv_1}(\omega, \mathbf{R}_B, \mathbf{R}_A) d_{eg}^{*Av_1} d_{eg}^{Av_2} \delta(\omega - \omega_A) D_{amed}^{v_2 v'}(\omega, \mathbf{R}_B, \mathbf{R}_A) \right].$$

We should mention here that if both atoms are in their ground states, then instead of $\omega_A$ we should substitute $-\omega_A$. As a result, according to (63) the second term of (63) will be equal to zero. And we will arrive at the generalization of the result, obtained in [29] for interaction of two ground-state atoms in a vacuum. To obtain the result [29], we should take free photon Green functions and put $\gamma_B = 0$.

After averaging over all possible mutual orientations of dipole moments of atoms, we can write [29]

$$d_{eg}^{v_1} d_{ge}^{v_2} \to \frac{|d_{eg}|^2}{3} \delta_{v_1 v_2}, \quad \alpha^{v_1 v_2} \to \alpha \delta_{v_1 v_2} \tag{64}$$

Now, taking into account [29] $D_a(\omega) = D_r^*(\omega)$, we find

$$U(\mathbf{R}_A - \mathbf{R}_B) = \mathrm{Re}\frac{i|d_{eg}^B|^2}{6\pi} \int_0^\infty d\omega \left( \frac{1}{\omega_B - \omega - i\frac{\gamma_B}{2}} + \frac{1}{\omega_B + \omega - i\frac{\gamma_B}{2}} \right)$$
$$\times \sum_{v_1 v_2} \left[ \left( D_{rmed}(\omega, \mathbf{R}_B, \mathbf{R}_A) \right)^2 \alpha_A(\omega) + \frac{2\pi i}{3} \left| D_{rmed}(\omega, \mathbf{R}_B, \mathbf{R}_A) \right|^2 \left| d_{eg}^A \right|^2 \delta(\omega - \omega_A) \right]. \tag{65}$$

For the ground-state atom B the expression in parentheses resembles the polarizability of the ground-state atom (61). The difference is in the sign of $\gamma_B$ in the second – non-resonant term. For positive frequencies we can omit this difference and write

$$U(\mathbf{R}_A - \mathbf{R}_B) = \mathrm{Re}\frac{i}{2\pi}\left[ \int_0^\infty d\omega \alpha_B(\omega) \alpha_A(\omega) \left( D_{rmed}(\omega, \mathbf{R}_B, \mathbf{R}_A) \right)^2 \right.$$
$$\left. + \frac{2\pi i}{3} \int_0^\infty d\omega \alpha_B(\omega) \left| D_{rmed}(\omega, \mathbf{R}_B, \mathbf{R}_A) \right|^2 \left| d_{eg}^A \right|^2 \delta(\omega - \omega_A) \right]. \tag{66}$$

To find the result we should substitute (50) into (66)

$$U(R) = \text{Re} \frac{i}{9\pi} \left[ \int_0^\infty \alpha_{eA}(\omega) \alpha_{gB}(\omega) \frac{\omega^4}{R^2} \left( 1 + \frac{2i}{n(\omega)\omega R} - \frac{5}{(n(\omega)\omega R)^2} \right. \right.$$

$$\left. - \frac{6i}{(n(\omega)\omega R)^3} + \frac{3}{(n(\omega)\omega R)^4} \right) \exp(2in(\omega)\omega R) d\omega \tag{67}$$

$$+ 2\pi i |d_{eg}^A|^2 \alpha_{gB}(\omega_A) \frac{\omega_A^4}{R^2} \left( 1 + \frac{1}{(n(\omega_A)\omega_A R)^2} + \frac{3}{(n(\omega_A)\omega_A R)^4} \right) \exp\left[ -2\,\text{Im}(n(\omega_A))\omega_A R \right] \Bigg].$$

Here we use $R = R_A - R_B$.

Since polarizabilities of atoms have no poles in upper complex plane of $\omega$, the first term of (67) can be represented as an integral over imaginary frequency $iu$.

$$U(R) = \text{Re} \left[ -\frac{1}{\pi} \int_0^\infty \alpha_{eA}(iu) \alpha_{gB}(iu) \frac{u^4}{R^2} \right.$$

$$\times \left( 1 + \frac{2}{n(iu)uR} + \frac{5}{(n(iu)uR)^2} + \frac{6}{(n(iu)uR)^3} + \frac{3}{(n(iu)uR)^4} \right) \exp(-2n(iu)uR) du \tag{68}$$

$$\left. - \frac{4}{9} \frac{|d_{eg}^A|^2 |d_{eg}^B|^2 \omega_B \omega_A^4}{(\omega_B^2 - \omega_A^2 - i\gamma_B \omega_A) R^2} \left( 1 + \frac{1}{(n(\omega_A)\omega_A R)^2} + \frac{3}{(n(\omega_A)\omega_A R)^4} \right) \exp\left[ -2\,\text{Im}(n(\omega_A))\omega_A R \right] \right].$$

Expression (68) is the potential corresponding to interaction between excited atom A and ground-state atom B embedded in an absorbing dielectric medium described by complex refractive index $n$. We took into account possible absorption of photons by the dielectric medium, which is described by the imaginary part of the refractive index. The first term is non-resonant. It corresponds to either interaction between excited and ground-state atoms, or interaction between two ground-state atoms. The second term is resonant. It describes interaction between excited and ground-state atoms only. If both atoms are not excited, this term is equal to zero. This term may result in either attraction, or repulsion between the atoms. If $\omega_A > \omega_B$ the interaction is repulsive, if $\omega_A < \omega_B$ the interaction is attractive. Besides, the resonance term may be up to four orders of magnitude greater than the non-resonance term. Niemax has measured the force between excited and ground-state atoms and confirmed the resonance repulsion between them [13,14].

Let us compare the result of (68) with the one obtained with the help of perturbation theory by Power and Thirunamachandran (1). Let $n=1$, $\gamma_B = 0$. In this case the formulae (68) and (1) coincide. The main difference of the result (68) from the one, obtained with the help of perturbation technique is the exponential factor in the second term $\exp\left[-2\,\text{Im}(n(\omega_A))\omega_A R\right]$. This factor results in the suppression of the van der Waals –Casimir-Polder interaction between excited and ground-state atoms in an absorbing dielectric medium. The reason for this suppression is the possibility for photons to be absorbed by the medium. Such absorption can not be taken into account in perturbation technique. As it will be shown in the next chapter, this suppression is crucial for Casimir-Polder limit of retardation ($R \gg \lambda$). As it is shown in the next chapter, this exponential factor makes the divergent integral (5) convergent.

Let us consider interaction between two ground-state atoms. We should substitute $\omega_A \to -\omega_A$. It results in the formula

$$U(R) = \text{Re}\left[-\frac{1}{\pi}\int_0^\infty \alpha_{gA}(iu)\alpha_{gB}(iu)\frac{u^4}{R^2}\right.$$
$$\left.\times\left(1+\frac{2}{n(iu)uR}+\frac{5}{(n(iu)uR)^2}+\frac{6}{(n(iu)uR)^3}+\frac{3}{(n(iu)uR)^4}\right)\exp(-2n(iu)uR)\,du\right] \quad (69)$$

For the non-retarded van der Waals limit ($R \ll \lambda$) the expression (69) reads

$$U(R) = -\text{Re}\left[\frac{3}{\pi}\int_0^\infty \frac{\alpha_{gA}(iu)\alpha_{gB}(iu)}{R^6(n(iu))^4}du\right] \quad (70)$$

For the retarded limit ($R \gg \lambda$)

$$U(R) = -\frac{23\alpha_{gA}(0)\alpha_{gB}(0)}{4\pi(n(0))^5 R^7}. \quad (71)$$

The expressions (70) and (71) coincide with the corresponding formulae, obtained for interaction between two ground-state atoms embedded in a dielectric medium[32].

# III. Interaction between an excited atom and a plane dielectric medium of ground-state atoms

We consider interaction between an excited atom and a dielectric medium of ground-state atoms. Let the medium be diluted so we can take into account only pair interactions. Let excited atom A be situated at a distance of $z_0$ from the interface of the medium of atoms B (Fig.4). Atoms A and B are dissimilar. The density number of the atoms of the medium is $n_0$. To find the interaction potential we should integrate expression (68) over the volume of the medium.

$$U_z(z_0) = \int dV U(R) n_0$$

This integral is convergent due to the exponential factor of the second term. If we take the expression obtained with the help of perturbation technique(1), the result (5) is divergent.

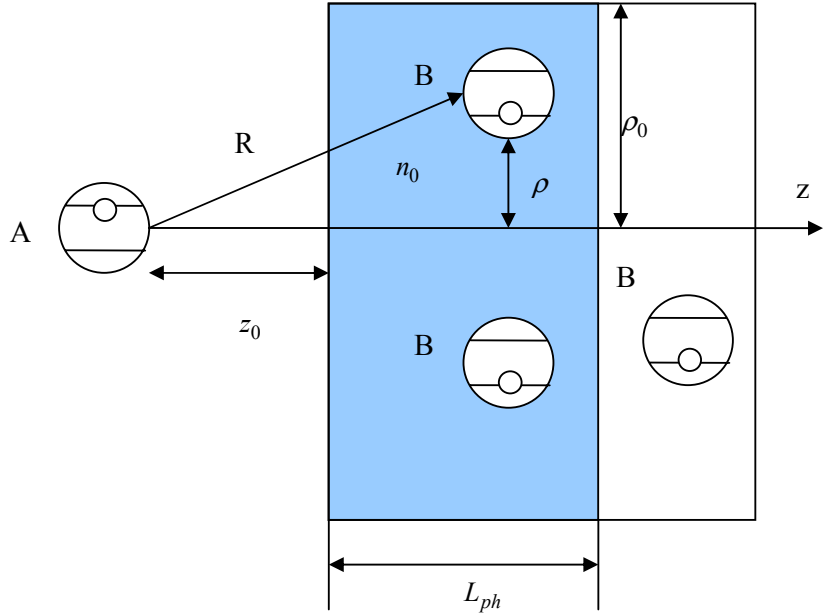

**Fig.4 Interaction between excited atom A and ground-state atoms B**

To simplify calculations, we will use the following model. Let

$$L_{ph} = \left(2 \operatorname{Im}(n(\omega_A)) \omega_A\right)^{-1} \tag{72}$$

be the photon free mean path in the medium. We will restrict the volume of integration by the photon free mean path and drop the exponent in the second term of the expression(68). The resonance part of the force between two atoms in this model is

$$F_R(R) = \frac{\partial}{\partial R}U(R) = \text{Re}\left[\frac{8}{9}\frac{\left|d_{eg}^A\right|^2\left|d_{eg}^B\right|^2\omega_B\omega_A^4}{\left(\omega_B^2-\omega_A^2-i\gamma_B\omega_A\right)R^3}\left(1+\frac{2}{(\omega_A R)^2}+\frac{9}{(\omega_A R)^4}\right)\right]. \quad (73)$$

Here we put $n \approx 1$ in denominators of (73) for dilute gas.

Integrating the force (73) over the volume shown in Fig.4 gives us

$$F_z(z_0) = \text{Re}\frac{4\pi}{9}\frac{\left|d_{eg}^A\right|^2\left|d_{eg}^B\right|^2\omega_B\omega_A^4 n_0}{\left(\omega_B^2-\omega_A^2-i\gamma_B\omega_A\right)}\left[\ln\left(1+\frac{L_{ph}}{z_0}\right)\right.$$
$$\left.+\frac{2}{\omega_A^2}\left(\frac{1}{z_0^2}-\frac{1}{\left(z_0+L_{ph}\right)^2}\right)+\frac{3}{2\omega_A^4}\left(\frac{1}{z_0^4}-\frac{1}{\left(z_0+L_{ph}\right)^4}\right)\right]. \quad (74)$$

For dilute gas the permittivity is

$$\varepsilon(\omega) = 1+4\pi n_0 \alpha_g(\omega).$$

Substituting here the polarizability of atom B (61) with (64), we find

$$n(\omega) = \sqrt{\varepsilon(\omega)} = \sqrt{1+\frac{4\pi}{3}n_0\left|d_{ge}^B\right|^2\left(\omega_B^2-\omega^2-i\gamma_B\omega\right)^{-1}}$$

Using expression (72) we find

$$L_{ph} = 3\frac{\left(\omega_B^2-\omega_A^2\right)^2+\left(\gamma_B\omega_A\right)^2}{4\pi n_0\left|d_{ge}^B\right|^2\gamma_B\omega_A^2}. \quad (75)$$

Let $z_0 \gg L_{ph}$, then

$$F_z(z_0) = \frac{\left|d_{eg}^A\right|^2\omega_B\omega_A^2\left(\omega_B^2-\omega_A^2\right)}{3z_0\gamma_B}.$$

We should stress here that the force does not depend on the density number of the medium. The force between an excited atom and a gaseous medium is not oscillating as it was expected for metal [19]. For the opposite case $z_0 \ll L_{ph}$ and $z_0 \ll \lambda$, we obtain

$$F_z(z_0) = \frac{2\pi}{3}\frac{\left|d_{eg}^A\right|^2\left|d_{eg}^B\right|^2\omega_B n_0\left(\omega_B^2-\omega_A^2\right)}{\left(\left(\omega_B^2-\omega_A^2\right)^2+\left(\gamma_B\omega_A\right)^2\right)z_0^4}. \quad (76)$$

This result coincides with the ones, obtained for interaction of excited atom and a dielectric medium for the case of dilute gas medium and $z_0 \ll \lambda$ [19][20]. But here, we introduce another parameter – the free mean path of the photon in the medium. Our formula (74) coincides with the previous results [19][20] for dilute gas only if $z_0 \ll L_{ph}$.

## IV. Excited atom in the center of a hemisphere of dilute gas medium

Let excited atom A be situated in the center of a hemisphere of a gas of ground-state atoms B (Fig.5)

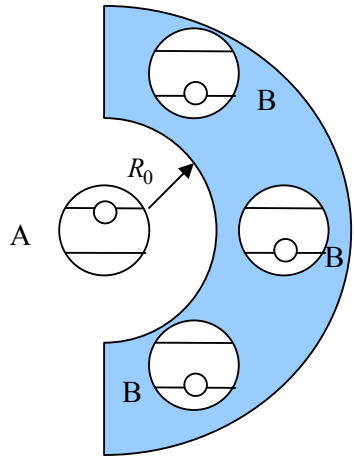

**Fig.5 Interaction between an excited atom and a hemisphere**

The resonance part of the force of interaction between a single atom B and atom A can be found using (68) as follows

$$F_R(R) = \frac{\partial}{\partial R} U(R) = \text{Re} \frac{4}{9} \frac{\left|d_{eg}^A\right|^2 \left|d_{eg}^B\right|^2 \omega_B \omega_A^4}{\left(\omega_B^2 - \omega_A^2 - i\gamma_B \omega_A\right) R^2} \left[ \frac{2}{R}\left(1 + \frac{2}{\omega_A^2 R^2} + \frac{9}{\omega_A^4 R^4}\right) \right.$$
$$\left. + \frac{1}{L_{ph}}\left(1 + \frac{1}{\omega_A^2 R^2} + \frac{3}{\omega_A^4 R^4}\right) \right] \exp\left(-(R-R_0)/L_{ph}\right). \qquad (77)$$

After integration of (77) over the infinite volume of the hemisphere, we arrive at

$$F(R_0) = \frac{4\pi}{9} \frac{\left|d_{eg}^A\right|^2 \left|d_{eg}^B\right|^2 \omega_B \omega_A^4 \left(\omega_B^2 - \omega_A^2\right) n_0}{\left(\left(\omega_B^2 - \omega_A^2\right)^2 + \left(\gamma_B \omega_A\right)^2\right)} \left[1 + 2\frac{L_{ph}}{R_0}\right] \qquad (78)$$

for $R_0 \gg L_{ph}$, $R_0 \gg \lambda$, and

$$F(R_0) = 2\pi \frac{|d_{eg}^A|^2 |d_{eg}^B|^2 \omega_B \omega_A^4 \left(\omega_B^2 - \omega_A^2\right)^2 n_0}{\left(\left(\omega_B^2 - \omega_A^2\right)^2 + (\gamma_B \omega_A)^2\right) R_0^4}, \tag{79}$$

for $R_0 \ll L_{ph}$, $R_0 \ll \lambda$.

## VI. Summary

The retarded interaction between two dissimilar atoms if one of them is excited is described by the interaction potential, which is proportional to $R^{-2}$. This result was obtained with the help of perturbation technique [8,9]. So, if we consider interaction between an excited atom and a dilute gas of ground-state atoms using the lowest orders of perturbation technique, we obtain divergence.

To solve the problem of divergence, we considered interaction between two atoms (one of them is excited) embedded in an absorbing dielectric medium. To calculate the interaction, we used the method based on the kinetic Green function [24]. We took into account absorption of photons in the medium. So, the resonance part of the potential is proportional to $\exp\left[-2\operatorname{Im}(n(\omega_A))\omega_A R\right]$. It results in the suppression of the Casimir-polder interaction between excited and ground-state atoms. Besides, we took into account finite width of the excited level of one of the atoms. In the limiting case $n \to 1$ and $\gamma \to 0$ we obtained the result of perturbation theory [8,9]. If the atoms are in their ground state, our result coincides with the one obtained by S. Buhmann et al. [32] for two ground-state atoms embedded in a dielectric medium.

Then we considered interaction between an excited atom and a gas medium of ground-state atoms. We took into account only pair interactions. The result is no more divergent. We introduced a new parameter – the free mean path of the photon in the absorbing medium $L_{ph}$. We showed that for $z_0 \ll L_{ph}$ and $z_0 \ll \lambda$, the conventional result for non-retarded interaction between an excited atom and a medium for a case of dilute gas [19,20] is obtained. But for $z_0 \gg L_{ph}$ and $z_0 \ll \lambda$ the result is different despite the non-retarded character of the interaction. We obtained the Casimir-Polder force for $z_0 \gg L_{ph}$ and $z_0 \gg \lambda$. It does not depend on the density number of the medium. But we should stress here, that for limiting case

$n_0 = 0$ the inequality $z_0 \gg L_{ph}$ is violated. The obtained results are of resonance character, but for equal transition frequencies of the atoms it can not be applied. For the case of identical atoms one should describe the atoms by a single symmetric or antisymmetric wave function [10].

We calculated the force on an excited atom situated in the center of a hemisphere of dilute gas of ground-state atoms. We obtained an exact solution of the problem.

## Appendix A. Derivation of equation (10)

Let us consider interaction between two atoms placed at points with radius-vectors $\boldsymbol{R}_A$ and $\boldsymbol{R}_B$. For the sake of simplicity we consider hydrogen-like atoms here. Let the interaction potential be $U(\boldsymbol{R}_A - \boldsymbol{R}_B)$. The masses of the atoms are large enough to neglect the motion of the nuclei. Thus, the Shrodinger equation for this system reads

$$\left(\hat{H}_A + \hat{H}_B + U(\boldsymbol{R}_A - \boldsymbol{R}_B)\right)\Psi = E\Psi, \tag{80}$$

where $\hat{H}_A = -\dfrac{\Delta_{r_A}}{2m_e} + U_A(\boldsymbol{r}_A - \boldsymbol{R}_A)$ and $\hat{H}_B = -\dfrac{\Delta_{r_B}}{2m_e} + U_B(\boldsymbol{r}_B - \boldsymbol{R}_B)$ are the Hamiltonians of free atoms A and B, $m_e$ is the mass of the electron. If the distance between the atoms is large to neglect the exchange interaction (we consider long-range interaction), and the atoms are dissimilar to neglect the resonance interaction between them, we can represent the eigen-function of the system as follows

$$\Psi = \Psi_A(\boldsymbol{r}_A - \boldsymbol{R}_A)\Psi_B(\boldsymbol{r}_B - \boldsymbol{R}_B).$$

Now we can apply the Hartree-Fock method[33] to Eq.(80).

$$\left(\hat{H}_A + V_A\right)\Psi_A = E_A \Psi_A, \tag{81}$$

$$\left(\hat{H}_B + V_B\right)\Psi_B = E_B \Psi_B, \tag{82}$$

where

$$V_A = \int \Psi_B^*(\boldsymbol{r}_B - \boldsymbol{R}_B) U(\boldsymbol{R}_A - \boldsymbol{R}_B) \Psi_B(\boldsymbol{r}_B - \boldsymbol{R}_B) d\boldsymbol{r}_B = U(\boldsymbol{R}_A - \boldsymbol{R}_B) \tag{83}$$

and

$$V_B = \int \Psi_A^*(r_A - R_A) U(R_A - R_B) \Psi_A(r_A - R_A) dr_A = U(R_A - R_B). \tag{84}$$

Using equations (83) and (84), we can rewrite Eqs (81) and (82)

$$\hat{H}_A \Psi_A = (E_A - U(R_A - R_B)) \Psi_A,$$

$$\hat{H}_B \Psi_B = (E_B - U(R_A - R_B)) \Psi_B.$$

Consequently, $U(R_A - R_B) = \Delta E_A = \Delta E_B$.

Strictly speaking, the Hartree-Fock method is applicable to the interaction between ground-state systems, but here we suppose that the excited state is a long-living one (the life-time is supposed to be infinite). So, we can apply this method to our problem.

## Appendix B Derivation of the equations (26) and (27)

To obtain the Eqs (26) and (27) we have to expand expression (16) and substitute the result into (17). Then, using (13) and Wick's theorem, we find

$$\begin{aligned}
\rho_{1c}^B(x, x') = -i \int dx_1 dx_2 g_{11}^{0B}(x, x_1) \hat{d}^\nu \hat{d}^{\nu'} g_{11}^{0B}(x_1, x_2) \\
\times \Bigg\{ D_{11}^{0\nu\nu'}(x_2, x_1) + \int dx_3 dx_4 \sum_{\nu_1 \nu_2} \Big[ D_{11}^{0\nu\nu_1}(x_2, x_3) \Pi_{11}^{\nu_1\nu_2}(x_3, x_4) D_{11}^{0\nu_2\nu'}(x_4, x_1) \\
+ D_{11}^{0\nu\nu_1}(x_2, x_3) \Pi_{12}^{\nu_1\nu_2}(x_3, x_4) D_{21}^{0\nu_2\nu'}(x_4, x_1) \\
+ D_{12}^{0\nu\nu_1}(x_2, x_3) \Pi_{21}^{\nu_1\nu_2}(x_3, x_4) D_{11}^{0\nu_2\nu'}(x_4, x_1) + D_{12}^{0\nu\nu_1}(x_2, x_3) \Pi_{22}^{\nu_1\nu_2}(x_3, x_4) D_{21}^{0\nu_2\nu'}(x_4, x_1) \Big] \Bigg\} \rho_0^B(x_2, x'). 
\end{aligned} \tag{85}$$

Here, to simplify the equations we consider only the term represented by the Feynman's diagram in Fig.1c and the terms of higher orders of the same topology. Other terms given in Fig.1 can be treated the same way.

The polarization operators of the system are

$$\Pi_{11}^{\nu_1\nu_2}(x_1,x_2) = -\hat{d}^{\nu_1}\hat{d}^{\nu_2}\left(\rho^A(x_1,x_2)g_{11}^A(x_2,x_1) + g_{11}^A(x_1,x_2)\rho^A(x_2,x_1)\right) + \Pi_{11med}^{\nu_1\nu_2}(x_1,x_2),$$

$$\Pi_{12}^{\nu_1\nu_2}(x_1,x_2) = \hat{d}^{\nu_1}\hat{d}^{\nu_2}\left(\rho^A(x_1,x_2)g_{21}^A(x_2,x_1) + g_{12}^A(x_1,x_2)\rho^A(x_2,x_1)\right) + \Pi_{12med}^{\nu_1\nu_2}(x_1,x_2),$$

$$\Pi_{21}^{\nu_1\nu_2}(x_1,x_2) = \hat{d}^{\nu_1}\hat{d}^{\nu_2}\left(\rho^A(x_1,x_2)g_{12}^A(x_2,x_1) + g_{21}^A(x_1,x_2)\rho^A(x_2,x_1)\right) + \Pi_{21med}^{\nu_1\nu_2}(x_1,x_2),$$

$$\Pi_{22}^{\nu_1\nu_2}(x_1,x_2) = -\hat{d}^{\nu_1}\hat{d}^{\nu_2}\left(\rho^A(x_1,x_2)g_{22}^A(x_2,x_1) + g_{22}^A(x_1,x_2)\rho^A(x_2,x_1)\right) + \Pi_{22med}^{\nu_1\nu_2}(x_1,x_2),$$

(86)

where $\Pi_{med}^{\nu_1\nu_2}$ is the polarization operator of the medium. The first term of the expansion of the polarization operator for the medium is

$$\Pi_{11med}^{\nu_1\nu_2} = -\hat{d}^{\nu_1}\hat{d}^{\nu_2}\left(\rho^{med}(x_1,x_2)g_{11}^{med}(x_2,x_1) + g_{11}^{med}(x_1,x_2)\rho^{med}(x_2,x_1)\right). \tag{87}$$

The Eq.(87) corresponds to a single scattering of a photon on an atom of the medium. This polarization operator results from the part of the Hamiltonian (6) corresponding to the atoms of the medium.

To obtain the equation for the photon Green functions, one should substitutes Eq. (16) into (28) and use Wick's theorem. As a result, we obtain the expression in braces of Eq.(85). So, we find

$$\rho_{1c}^B(x,x') = -i\int dx_1 dx_2 g_{11}^{0B}(x,x_1)\hat{d}^\nu \hat{d}^{\nu'} g_{11}^{0B}(x_1,x_2) D_{11}^{\nu\nu'}(x_2,x_1)\rho_0^B(x_2,x').$$

Considering other terms given in Fig.1, we arrive at the Eq.(26) with the mass operators (27).

## Acknowledgement

I would like to thank Prof. B.A. Veklenko for permanent attention to this work and Stefan Buhmann for valuable discussion and [32].